\documentclass[twocolumn,aps,prb,letterpaper]{revtex4}

\usepackage{epsfig}
\usepackage{epsf}
\usepackage{array}
\usepackage{amssymb}
\usepackage{amsbsy}
\usepackage{amsmath}
\usepackage{revsymb}

\newcommand{\be}{\begin{equation}}
\newcommand{\ee}{\end{equation}}
\newcommand{\bea}{\begin{eqnarray}}
\newcommand{\eea}{\end{eqnarray}}
\newcommand{\w}{\omega}
\newcommand{\W}{\Omega}
\newcommand{\e}{\epsilon}
\newcommand{\q}{\boldsymbol{q}}

\renewcommand{\Im}{\mathrm{Im}}

\newcommand{\req}[1]{Eq.~(\ref{#1})}
\newcommand{\reqs}[1]{Eqs.~(\ref{#1})}
\newcommand{\rref}[1]{(\ref{#1})}

\newcommand{\Ha}{{\cal H}}

\newcommand{\ocite}[1]{Ref.~\onlinecite{#1}}

\begin{document}

\title{Tunneling anomaly of superconducting films in strong magnetic fields.}

\author{G. Catelani}
\altaffiliation[Current address: ]{Laboratory of Atomic and Solid State 
Physics, Cornell University, Ithaca, NY 14853}
\affiliation{Physics Department, Columbia University, New
York, NY 10027}

\begin{abstract}
We consider the tunneling Density of States (DoS) of superconducting films 
driven to the paramagnetic phase by the Zeeman splitting. 
We show that there is minimum in the DoS whose position depends on the 
orientation of the applied field. This dependence, not predicted by previous 
theoretical calculations, is in agreement with a recent experiment.
\end{abstract}

\date{\today}
\maketitle

It is well known that superconductivity can be destroyed by applying a  
magnetic field because of the breaking of the time reversal symmetry
(see e.g. \ocite{tb}). The magnetic field acts on both the orbital motion
and the spin of the electrons; while in the bulk the orbital effect 
is responsible for the suppression of superconductivity, in low-dimensionality 
systems ($d\leq 2$) the Zeeman splitting of the 
states with opposite spin and which are degenerate in zero field
can be the dominant mechanism for this suppression.\cite{clch,igor}
In the normal phase, the theory of interaction corrections to the Density of 
States (DoS) $\nu$  reviewed in \ocite{AA} predicts the appearance of singular
contributions to the DoS located at the Zeeman energy $E_Z= g_L \mu_B H$
($g_L$ is the Land\'e g-factor, $\mu_B$ the Bohr magneton and $H$ the magnetic
field). However  
experiments performed almost a decade ago\cite{ad1} and subsequent theoretical 
calculations\cite{igor} showed that there are more singular corrections located
at a lower energy $E_+$:
\be\label{ep}
E_+ = \left( E_Z + \Omega \right)/2 \, ,
\ee
where
\be\label{om}
\Omega=\sqrt{E_Z^2-\Delta^2}
\ee
and $\Delta$ is the BCS gap. In a small grain, the origin of these corrections 
can be understood as follows:\cite{igor} when a 
spin-down electron tunnels into a state singly occupied by a spin-up electron, 
they form a pair; the interaction mixes this pair with the empty states at 
energies $> E_Z/2$; the mixing is resonant at the energy $E_+$ and this 
resonance manifests itself as an anomalous contribution to the DoS.
The position of the anomaly was predicted to be 
``universal'', i.e. independent of both the dimensionality and the direction 
of the magnetic field. Recent experiments on superconducting Al 
films\cite{ad2} are in disagreement with the latter result: the position of the
measured minimum in the DoS moves to higher energies as the component of the 
magnetic field perpendicular to the film is increased. 

In this paper we reconsider the calculation of 
the superconducting fluctuations corrections to the DoS in the normal phase
for disordered films and wires in strong magnetic fields. 
To understand why the reconsideration is necessary, let us briefly review
the qualitative argument given in Ref.~\onlinecite{igor} to explain the
position and width of the anomaly in films and wires: as discussed above, 
the position of the singularity was found to be located at $E_+$ for tunneling 
into a grain; then a self-consistent argument was given to find the 
characteristic energy scale $W_d$ (width of the singularity in $d$ dimensions) 
in one- and two-dimensional systems. The latter argument is based on the 
assumption that the system can be effectively divided into zero-dimensional 
patches whose size $L_{W_{d}}$ is then 
found self-consistently -- this assumption however breaks down if dephasing 
happens on a scale shorter than the patches' size. In the presence of a 
perpendicular magnetic field, this scale is given for films by the magnetic 
length $l_H=(\hbar c/eH)^{1/2}$, and the break-down happens at
\be\label{cond2}
l_H \sim L_{W_{2}}
\ee
where $L_{W_{d}}$ is:\cite{igor}
\be
L_{W_{d}}= \left( \frac{\hbar D}{W_d} \right)^{1/2} .
\ee
This suggests that, for strong enough magnetic fields, 
additional contributions 
to the DoS which are particular to one- and two-dimensional systems might
become relevant;\footnote{A similar argument, valid for 
wires and films in parallel field, is obtained by substituting $l_H$ with 
$L_H\equiv \sqrt{D\tau_H}$, with $\tau_H$ defined in 
\req{thdef}.}
 below we show that this is indeed the case and 
for films in a tilted field, we present the detailed derivation of the 
dependence of the position of the singularity on the 
perpendicular component of the  magnetic field.
From now on, we use units with $\hbar=1$.

To describe the superconducting systems, we consider the pairing Hamiltonian 
$\Ha=\Ha_0+\Ha_{int}$, where the non-interacting part $\Ha_0$ is given by the 
sum of the kinetic energy, the Zeeman energy and the disorder potential, and 
$\Ha_{int}$ is, in second-quantized notation ($g>0$ is the coupling constant):
\[
\Ha_{int} = -g\int\!d^dr \, \psi^\dag_\uparrow(r)\psi^\dag_\downarrow(r) 
\psi_\downarrow(r) \psi_\uparrow(r) \, .
\]
The DoS is related to the imaginary part of the one particle Green's function
and the latter can be calculated with e.g. the diagrammatic 
technique.\cite{agd} As this procedure is standard, we skip intermediate steps 
and we simply quote the 
final answer\cite{AA} for the one-loop fluctuation correction to the DoS for 
spin down electrons at zero temperature in $d$-dimensions:
\be\label{dnform}
\frac{\delta \nu_d (\e)}{\nu_0} = -\frac{1}{\pi} \Im
\int_{-\infty}^{\ \e}\!\!\!\!\!d\w\!\int\!\!\frac{d^dq}{(2\pi)^d}\,
\Lambda(|\w|,\q) C^2(2\e -\w-E_Z,\q)
\ee
where $\nu_0$ is bare DoS per spin, 
$C$ is the Cooperon:\footnote{The expressions \rref{coopdef} and 
\rref{lamdef} are in the diffusive approximation, valid for $T\tau\ll 1$ with
$T$ the temperature and $1/\tau$ the impurity scattering rate.}
\be\label{coopdef}
C(\e,\q) = \frac{1}{-i\e + Dq^2}
\ee
with $D$ the diffusion constant,
and $\Lambda$ is the fluctuation propagator:
\be\label{lamdef}
\Lambda(\w,\q) = \frac{2}{\nu_0} \left[ \ln \left(
\frac{E_Z^2 + (-i\w +Dq^2)^2}{\Delta^2}\right) \right]^{-1}  \, .
\ee
As noticed in Ref.~\onlinecite{igor}, due to its analytical properties this 
propagator can be separated into a ``pole'' part and a ``cut'' part:
\be\label{sep}
\Lambda = \Lambda^{p} + \Lambda^c \, ,
\ee
where
\be\label{lpol}
\Lambda^{p} (\w,\q) \equiv \frac{2}{\nu_0} \frac{\Delta^2}{2\W} 
\frac{i}{-i(\w-\W)+Dq^2}
\ee
and $\Lambda^c$ is implicitly defined by \reqs{sep}-\rref{lpol}. We can 
accordingly write $\delta\nu_d$ as a sum of two terms:
$\delta\nu_d (\e) = \delta\nu^p_d (\e) + \delta\nu^c_d(\e)$.
The contribution $\delta\nu^c_d(\e)$ can be found 
in Ref.~\onlinecite{AA} and for example in $d=2$ it is proportional to 
$\ln [\ln (\e -E_Z)]$; we give it here no further consideration,
since this contribution is less divergent than the ones we 
calculate below. We note that the separation \rref{sep} is possible provided
that $E_Z > \Delta$; if this condition is satisfied, our results
are applicable even for fields smaller than the parallel critical field 
$H_{c\parallel}$ as
long as the sample is in the normal state.

\begin{figure}[!tb]
\begin{center}\includegraphics[width=0.46\textwidth]{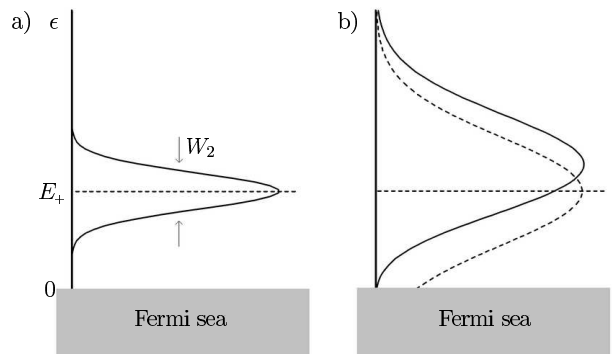}\end{center}
\caption{Width and position of the minimum for small (left) and large (right)
perpendicular field as explained in the text.}
\label{fig1}
\end{figure}

According to the above argument, we want to evaluate the right hand side of
\req{dnform} with the substitution 
\be\label{subst}
\Lambda \to \Lambda^p 
\ee
and the result of this calculation gives the lowest order perturbative 
correction to the DoS. On the other hand in \ocite{igor} a resummation of 
perturbation theory was performed in the long wavelength approximation 
$D q^2 \lesssim \Delta$, but this approximation is not always
applicable. Indeed let us consider a 2D system, in which case the approximation
means that, in energy units, the exchanged momentum is limited by  
$\Delta/g$, where $g=4\pi \nu_0 D$ is the adimensional conductance;
for good conductors $\Delta/g \ll \Delta$ and this
energy scale is much smaller than the gap $\Delta$.
Since the Cooper pair energy [i.e. the position of the pole in \req{lpol}]
$\Omega$ is $\gtrsim \Delta$
and the exchanged energy is much smaller than $\Omega$, we can effectively
neglect the ``Fermi sea'' under the pair: from a formal point of view, we 
can extend the upper limit of integration in \req{dnform} to infinity.
In other words, in this approximation we can neglect the exclusion principle, 
which forbids the 
electrons from ``going under the sea'', i.e. interacting with electrons at 
energies below the Fermi energy.\footnote{This is the more rigorous
justification of the ``frozen electrons'' 
assumption used in the qualitative discussion (Section IV) of 
Ref.~\onlinecite{igor}.}  
However when a magnetic field with a component perpendicular to the film is 
present, Landau quantization renders the momentum variable discrete:
\be\label{perfs1}
Dq^2 \to \Omega_H \left( n +\frac{1}{2} \right) \, ,
\ee
where
\be\label{wh}
\Omega_H = 4eDH \sin \theta / c
\ee
is the Cooperon cyclotron frequency and $\theta$ is the tilting angle.
The momentum transfer (in energy units) is therefore 
of order $\Omega_H$ and when this becomes sufficiently large we are not
allowed to neglect the exclusion principle anymore: the correct limits of 
integration must be considered.\footnote{The argument holds in the 
parallel field case and for wires if we substitute $\Omega_H$ with
$1/\tau_H$.} For strong fields, the effect of the 
``Fermi sea'' is to move the position of the minimum to higher 
energies:\footnote{These considerations, and Fig.~\ref{fig1}, were 
presented in \ocite{ad2}; they are reported here to make the paper 
self-contained.}
for $\Omega_H\ll W_2$ the anomaly, centered at $E_+$
has a width $W_2$ much smaller than $E_+$ [cf. \reqs{ep} and \rref{w2}] and
the Fermi sea is not probed 
by the excitations contributing to the anomaly, see Fig.~\ref{fig1}a.
On the contrary, for $\Omega_H\gg W_2$ a sizable fraction of these excitations 
would be located below the Fermi
energy, as shown by the dashed line in Fig.~\ref{fig1}b. However the exclusion 
principle suppresses the contributions from under 
the ``Fermi sea'', the profile of the anomaly becomes asymmetric and as a 
consequence the minimum appears to shift to higher energies, as in the solid 
line of Fig.~\ref{fig1}b. This qualitative argument predicts that  
as we increase $\Omega_H$, the shift also increases, in 
agreement with the quantitative result derived below. It also places the
transition between the weak and strong field regimes at $\Omega_H\sim W_2$,
which is the same condition as in \req{cond2}. In summary, the long wavelength
approximation is justified in the weak field regime and the non-perturbative
approach of \ocite{igor} is necessary in this case; in the strong field
limit, on the contrary, a perturbative calculation is sufficient, as we
explicitly show below.

After the substitution \rref{subst}, the integration over $\w$ in \req{dnform}
can be performed exactly and we write the result as:
\be
\delta\nu^p_d = \delta\nu_d^m+\delta\nu_d^l + \delta\nu_d^f \, ,
\ee
where we separated different contributions based on their degree of
divergence in the parallel field case: $\delta\nu_d^m$ gives the most divergent
term (a $\delta$-function in $d=2$) which was considered
in \ocite{igor}, along with subleading terms;  other subleading terms are 
collected in the the (less) divergent part $\delta\nu_d^l$;
finally $\delta\nu_d^f$ contains only finite contributions and hence we do not 
need its explicit form. 
The relevant contributions are:
\begin{subequations}\label{mycform}
\bea
\frac{\delta\nu_d^m(\e)}{\nu_0} &=& \frac{\Delta^2}{4\pi\nu_0 \W}\,\Im\!
\int\!\frac{d^dq}{(2\pi)^d}\,C^2(\e-E_+,\q) \label{mycform2} \\
&& \times \left[ \ln \frac{C(\e-E_Z,\q)}{C(\e-\W,\q)} 
- \ln\frac{C(2\e - E_Z,\q)}{C(-\W,\q)} \right] \nonumber 
\\
%\eea\bea
\frac{\delta\nu_d^l(\e)}{\nu_0} &=& \frac{\Delta^2}{2\pi\nu_0 \W}\,\Im\!
\int\!\frac{d^dq}{(2\pi)^d}\, C(\e-E_+,\q) \label{mycform1}\\ 
&& \quad \times \big[ C(\e-E_Z,\q) - C(2\e-E_Z,\q) \big] \nonumber \\ 
&& - \frac{i}{\e-E_-} \big[C(2\e-E_Z,\q) - C(-\W,\q) \big] \nonumber
\eea\end{subequations}
with $E_+$  and $\W$ defined in \reqs{ep} and \rref{om} respectively 
and we introduced:
\be\label{em}
E_- = \left( E_Z - \Omega \right)/2 \, .
\ee
Note that \req{mycform2} reduces to the perturbative result of \ocite{igor}
upon replacing the square bracket with $-2\pi i$. In what follows we
retain only the most divergent terms, as 
finite and weakly divergent contributions have already been discarded in 
performing the replacement \rref{subst}.

\begin{figure}[!bt]
\begin{center}\includegraphics[width=0.48\textwidth]{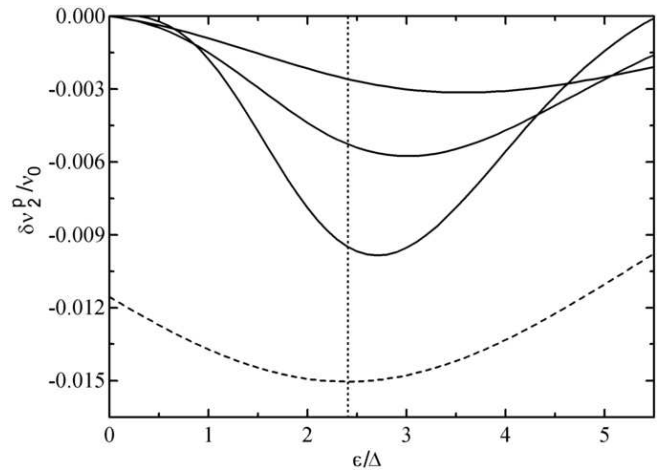}\end{center}
\caption{Tunneling anomaly in tilted field. The continuous lines are the  
correction $\delta\nu^p_2$ of \req{mycorr} for $\theta=35^\circ$, $23^\circ$ 
and $16^\circ$ (top to bottom) with $g=5$, $E_Z=2.5\Delta$ and $g_L=2$. 
The vertical dotted line is at $\e = E_+$.  The dashed line is the
approximate contribution of \ocite{igor} for $\theta=35^\circ$, which can
also be obtained by letting $b\to \infty$ in \req{fdef}.}
\label{fig2}
\end{figure}

We now restrict our attention to the case $d=2$ with a non-zero
perpendicular component of the magnetic field, so that we must use the 
substitution \rref{perfs1} and replace in \reqs{mycform} the integrations over 
momentum with summations:
\be\label{sumsub}
\int\!\frac{d^2q}{(2\pi)^2} \to \frac{\W_H}{4\pi D} \sum_{n=0}^{\infty} \, .
\ee
The result for the most divergent contribution is:
\be\label{mycorr}\begin{split}
\frac{\delta\nu^p_2(\e)}{\nu_0} = -\frac{W_2\pi^2}{\Omega_H} & \bigg[
f\left( \frac{\e-E_+}{\Omega_H} ; \frac{E_-}{\Omega_H} \right) \\ 
& + f\left( \frac{\e-E_+}{\Omega_H} ; \frac{\e -E_-}{\Omega_H} \right) 
\bigg] \, ,
\end{split} \ee
where the energy
\be\label{w2}
W_2 \equiv \Delta^2 /4g\Omega
\ee
characterizes the width of the tunneling anomaly in the parallel field and the 
function $f(a;b)$ is defined as:
\be\label{fdef}
f(a;b) = \Im \!\int^{\; b}_{-b} 
\frac{dt}{\pi^3 t} \left[ \psi'\left( \frac{1}{2} - it -ia \right) 
-\psi'\left( \frac{1}{2} -ia \right) \right] \! ,
\ee
where $\psi'$ is the derivative of the 
digamma function. The validity of \req{mycorr} is restricted to the strong 
field regime:
\be\label{limit}
\Omega_H \gg \pi^2 W_2
\ee
in which the correction is indeed smaller than the bare DoS. 
As an example, we plot both \req{mycorr} and the corresponding approximate
formula of \ocite{igor} in Fig.~\ref{fig2}. We note that, in agreement with 
our previous discussion: the minimum is shifted to a higher energy; the 
contribution at small energies (near the Fermi sea) is suppressed; the overall 
shape is asymmetric about the minimum, in qualitative agreement with the
experiments.\cite{ad1,ad2} In addition, the anomaly
is smaller than the prediction of the approximate formula -- this could be 
relevant in a quantitative comparison with experiments.

For the purpose of 
calculating the field-dependent position of the minimum, we can differentiate
\req{mycorr} and then expand the result to first order in 
$\delta = (\e - E_+)/\Omega_H$ by assuming $\delta \ll \frac{1}{2} $.
Performing this calculation we find:
\be\label{nuder}
\frac{\delta\nu_2^{p \, \prime}}{\nu_0} = \frac{W_2\pi^2}{\W_H^2}\Big[ 
\beta \delta  + \gamma \Big] \, ,
\ee
where 
\be\label{bdef}
\beta = - \left[f''\!\!\left(0;\frac{E_-}{\Omega_H}\right) 
+ f''\!\!\left(0;\frac{\Omega}{\Omega_H}\right) 
+\ddot{f} \left(0;\frac{\Omega}{\Omega_H}\right) \right] \lesssim 2\pi^2
\ee
with the prime denoting the derivative with respect to the first argument
and the dot with respect to the second argument of $f$, and
\be
\gamma = - \dot{f} \left(0;\frac{\Omega}{\Omega_H}\right) 
\ge \frac{2}{\pi^3} \psi''\!\left(\frac{1}{2}\right) \, .
\ee
While the above inequality is exact (and the equality is valid for 
$\Omega/\Omega_H=0$), the one in \req{bdef} should be understood
as follows: for 
$E_-,\Omega \gtrsim \Omega_H/3$ the approximate equality holds, while at lower
values of these parameters is the upper limit  that applies.
Setting \req{nuder} to 0 and using for an estimate the limiting values of 
$\beta$ and $\gamma$, we find 
for the position $E^*$ of the minimum:
\be\label{mineq}
E^* = E_+ + \alpha \W_H \, , \quad \alpha = 
-\frac{1}{\pi^5} \psi''\left(\frac{1}{2}\right) \simeq 0.055 \, ,
\ee
which generalizes the result of Ref.~\onlinecite{igor} and reduces to it 
for $\Omega_H =0$.\footnote{The validity of this formula is however restricted
by \req{limit} as well as by the condition $\delta\ll 1/2$.} 
The smallness of the constant $\alpha$ justifies {\it a posteriori} the 
expansion. In the experiment of \ocite{ad2}, \req{mineq} has been successfully 
tested; below we comment on the applicability of this equation for 
experimentally relevant values of the parameters and we
give more reliable estimates for the dependence of $E^*$ on $\Omega_H$.

According to the definitions \rref{ep}, \rref{om} and \rref{em},
the inequalities $E_- < \W \leq E_+ \leq E_Z$ hold;
in the experiments\cite{ad2} all these quantities are $\sim \Delta$. This means
that $W_2 \sim \Delta/4g \ll \Delta$ for good conductors. On the other hand
\be
\frac{\Omega_H}{E_Z} = \frac{4}{g_L} g \sin\theta \, ,
\ee
so that for ``large'' tilting angles we have $\Omega_H \gg E_Z$; thus we 
find that the condition 
\rref{limit} is experimentally satisfied. However the same reasoning
shows that the conditions for the applicability of the upper limit estimate in
\req{bdef} are easily violated. A more detailed study of the correction 
\rref{mycorr} shows that for the position $E^*$ of the minimum we can 
distinguish two regimes for different ranges of the parameter $\Omega_H/E_Z$, 
namely:
\begin{subequations}\label{esfin}\bea
E^* &=& E_+ + \alpha_1 \left(\frac{\Omega_H}{E_Z},\frac{E_Z}{\Delta}\right) 
\Omega_H  \
, \quad 1\lesssim \frac{\Omega_H}{E_Z} \lesssim 7 \qquad \ \label{eslow} \\
E^* &=& 0.543 E_+ +\alpha_\infty \Omega_H \ , \qquad \qquad
 \frac{\Omega_H}{E_Z} \gtrsim 7 \label{eshigh}
\eea\end{subequations}
with
\be
\alpha_\infty  \simeq 0.144 \, .
\ee
The coefficient $\alpha_1$ depends weakly on the field through $E_Z$;
for fields larger than about twice the parallel critical field this dependence
can be neglected\footnote{Its role is to increase the numerical
coefficient in the right hand side of \req{a1} at smaller fields.} and we find:
\be\label{a1}
\alpha_1 = \frac{3}{4} \frac{\W_H}{E_Z} \psi^{(2)}\left(\frac{1}{2}\right) / 
\psi^{(4)}\left(\frac{1}{2}\right)  \simeq 0.016 \frac{4}{g_L}
g \sin \theta
\ee
independent of the field.
The result \rref{eslow} is obtained by expanding $\beta$ [\req{bdef}] as a 
function of the quantities $E_-/\Omega_H$, $\Omega/\Omega_H$, while 
\req{eshigh} is found by taking the limit $g\to\infty$ in the derivative of 
\req{mycorr}: in this case the finite value of $\delta$ [defined before 
\req{nuder}] is calculated numerically 
and then we evaluate the coefficients of the first order 
corrections in $E_-/\Omega_H, \, \Omega/\Omega_H$ 
-- these are responsible for the 
suppression of the $E_+$ term. We stress that in the tilted field the linear 
dependence of the position of the minimum on the field is a robust prediction.
We also notice that in the allowed region of the parameters, the coefficient 
$\alpha_1$ in front of $\Omega_H$ varies between $0.016$ and 
$0.112$, i.e. it agrees with $\alpha$ of \req{mineq} within a factor of 3;
since it is also true that $\alpha_\infty/\alpha < 3$,
the estimate \req{mineq} is a correct order-of-magnitude one. Indeed
that equation has been successfully applied in a study of a $g=5.6$ sample 
for which\footnote{except at the highest angle $\theta = 90^\circ$.} 
$\W_H/E_Z \lesssim 8$ in \ocite{ad2}.\footnote{The condition 
\rref{limit} limits the applicability of our results to 
$\theta > 4^\circ$ in that experiment.} 
Experiments in higher conductance samples, where $\Omega_H/E_Z \gg 7$,
would enable to test \req{eshigh} directly: its simpler dependence on the 
physical parameters as compared to \req{eslow} would make possible a more
quantitative comparison between theory and experiments.

Up to now, we have not considered the phase relaxation effect of the 
(parallel component of the) magnetic field, since it can be usually neglected 
in the tilted field.\cite{igor} This effect is accounted for by 
shifting the frequency in the definition \rref{coopdef} of the Cooperon:
$\e \to \e + i/\tau_H$, with\cite{AA}
\be\label{thdef}
1/\tau_H = \Omega_{H_{\parallel}}^2/48E_T \, .
\ee
Here $\Omega_{H_{\parallel}}$ is as in \req{wh} but with $\cos\theta$ instead 
of $\sin\theta$ and the transverse Thouless energy is: $E_T = D/a^2$,
where $a$ is the film's width. In the parallel field, momentum integration in 
\reqs{mycform} is straightforward; we do not give here the explicit answers 
(due to space limitations), but the qualitative results are the same as for
the tilted field upon replacing $\Omega_H \to 1/\tau_H$.
Also, we note that \reqs{mycform}
can be used to evaluate the correction to the DoS of superconducting wires; 
for both wires and the parallel field case, the one-loop approximation is 
valid when $W_d \tau_H \ll 1$.\footnote{The definition of 
$\tau_H$ in $d=1$ is as in \req{thdef} up to numerical coefficients; $a$ is 
then the width of the wire.}

In conclusion, we calculated the one-loop interaction correction to the Density
of States for superconducting films in tilted field and in the paramagnetic 
phase. We found a minimum in the Density of 
States which shifts linearly with the applied field to higher energies,
see e.g. \req{mineq}; this dependence has been
demonstrated experimentally.\cite{ad2}

 We thank I. Aleiner for the numerous interesting conversations;
communications with P. Adams regarding the experimental results are 
gratefully acknowledged.

\end{document}